\documentclass[aps,prd,preprint,groupedaddress,showkeys]{revtex4-1}

\usepackage{graphicx}
\usepackage{amsmath}
\usepackage{color}

\begin{document}


\title{Topological correlations and breaking of fermionic antisymmetry of electrons in FQHE}


\author{Janusz Jacak}
\email[]{janusz.jacak@pwr.edu.pl}
\affiliation{Department of Quantum Technologies, Faculty of Fundamental Problems of Technology, Wroc{\l}aw University of Science and Technology, Wyb. Wyspia\'nskiego 27, 50-370 Wroc{\l}aw, Poland}


\date{\today}

\begin{abstract}
Highly nonlocal interparticle correlations in quantum Hall states of 2D charged system exposed to the perpendicular strong magnetic field are detailed by application of the commensurability condition upon path-integral quantization approach and examined by Monte-Carlo Metropolis simulations in perfect consistence with exact diagonalization of the Coulomb interaction in small models and with experimental data. In this way refined filling rate hierarchy in the lowest Landau level fully explains the experimentally collected features for FQHE at filling ratios predicted by the conventional composite fermion model as well as for those beyond the composite fermion model but also visible in the experiment. The trial wave functions for FQHE states are proposed using a systematic topological method revealing the different symmetry for different correlated states depending on filling fraction. A violation of fermionic antisymmetry for 2D electrons is evidenced for some filling rates related to specific correlations in FQHE.
\end{abstract}


\maketitle

\section{Introduction}

Since the discovery of the fractional quantum Hall effect (FQHE) \cite{tsui1982} and its description by the famous Laughlin function \cite{laughlin2,laughlin1}, the problem of the long-range interparticle correlations on the plane between electrons exposed to the perpendicular quantizing strong magnetic field is still puzzling as manifesting itself at only specially chosen electron densities corresponding to the so-called FQHE hierarchy, whereas in close vicinity of these selected filling ratios correlations completely disappear. Experimentally noticed hierarchy of FQHE states embraces up to date ca. 70 various fractional fillings, majority of them in the lowest Landau level (LLL) including its spin polarization \cite{pan2003,jain-n}. The similar correlations at some selected fractional fillings are observed also in higher LLs, though the hierarchy from the LLL is not repeated there \cite{lls,ll8/3-1,5/2-1,ll8/3-2,xia2004}. The structure of the FQHE hierarchy is robust against a crystalline structure of the material which has been recently confirmed in graphene monolayer \cite{amet,dean2011,feldman,feldman2} and in bilayer graphene \cite{bil,kou2014science,maher2014science,kim2015nanolett,diankov2016naturecomm}, though in graphene the different Landau level quantization is induced by pseudo-relativistic Dirac-like electron dynamics \cite{geb}. 

The correlations in FQHE in the LLL duplicate correlations at the completely filled LLs corresponding to IQHE at filling fraction $\nu=1,2,\dots$, which manifests itself in vanishing of the longitudinal resistivity $R_{xx}$ and quantization of the Hall conductivity $\sigma_{yx}=\frac{j_x}{U_y}=\frac{e^2 \nu}{h}$, as visible in experiments \cite{pan2003}. The closeness between correlations corresponding to IQHE and FQHE is phenomenologically addressed upon the concept of composite fermions (CFs) \cite{jain} to mapping of FQHE at fractional fillings of the LLL onto IQHE of completely filled consecutive LLs in reduced magnetic field screened by the mean field of auxiliary fluxes pinned to composite fermions \cite{jain2007}. In this way the main line of CF hierarchy is established, $\nu=\frac{n}{(q-1)n\pm 1}$ ($n=1,2,3,\dots$, $q$-odd integer).

\begin{figure}[ht]
\centering
\resizebox{0.45\textwidth}{!}{\includegraphics{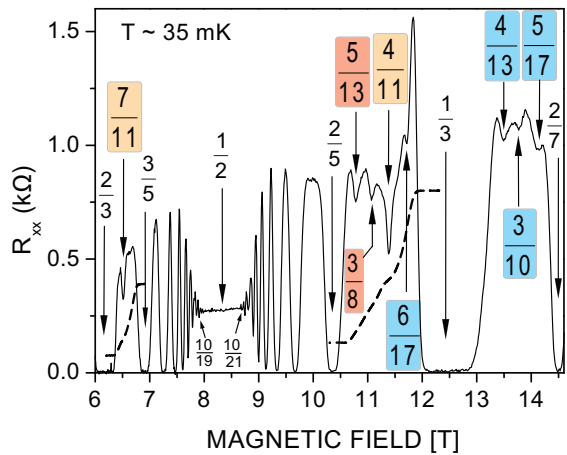}}
\caption{\label{fig2} Measured $R_{xx}$ in the fragment of the LLL in GaAs 2DEG (after \cite{pan2003}), in color are indicated fractions at which the similar level of $R_{xx}$ is achieved--- its nonzero value indicates that not all electrons are involved in the correlated state, as for correlations given by (\ref{444}) with $x>1$}
\end{figure}

Nevertheless, the full insight into correlations in FQHE is still incomplete as there are observed many fractions out of the main line of CF hierarchy (e.g., $\nu=\frac{4}{11}, \frac{5}{13}, \frac{3}{10}, \frac{3}{8}, \frac{5}{17}, \frac{6}{17}, \frac{7}{11}, \frac{4}{11}$, $\dots$) \cite{pan2003}. At theses fractions $R_{xx}$ typically is reduced in comparison to insulating phase but does not vanish which is in opposition to vanishing of $R_{xx}$ at filling fractions from the main CF series (cf. Fig. \ref{fig2}). This corresponds apparently to distinct type of correlations at these extraordinary filling fractions. 
They cannot be referred to higher LL correlations of CFs in resultant magnetic field screened by the averaged field of flux tubes pinned to electrons as it has been argued for the main line of CFs. Instead, an attempt to explain $\frac{4}{11}$ and $\frac{5}{13}$ states as the secondary generation of FQHE of the primary generation CFs is proposed in the paper \cite{jain2014}, which resolves itself to hypothetical dressing with interaction of already dressed CFs. This needs, however, the vague assumption that the residual interaction of primary CFs is still strong at some selected filling fractions, whereas in very close vicinity of these filling rates this additional interaction disappears. This is in conflict with an assertion \cite{jain2007} that CFs are some kind of quasiparticles which cannot change their residual interaction by a very delicate shift of the electron concentration. Similarly, some earlier study of the unconventional CF states out of the main series were not conclusive, cf. Refs \cite{d1,d2,d3}. These troubles of the theory indicate that the structure of FQHE correlations goes beyond the concept of CFs efficient only for the main CF series. In the present paper we clarify this situation and explain the correlations of FQHE in topological terms both for CFs and beyond, in fully consistence with the experimental observations \cite{d4,d5,pan2003}. 

The CF fermion structure has been referred to dressing of ordinary electrons with Coulomb interaction in some analogy to quasiparticle formation frequent in solids \cite{jain2007}. The rigorous way to define quasiparticles (e.g., Landau quasiparticles in metals) is their definition as poles of the retarded single-particle Green function \cite{abrikosov1975} which is unavoidably linked with the requirement of the mass operator continuity \cite{abrikosov1975}. This condition is, however, not fulfilled for 2D Coulomb interacting electrons in strong magnetic field and called originally by Laughlin as the 'particle separation quantization' \cite{laughlin2,laughlin1} which precludes the rigorous definition of CF as quasiparticle.

The FQHE observation was more recently supplemented by rich experimental data in graphene \cite{dean2011,amet,feldman,feldman2} revealing the different hierarchy of FQHE in higher LLs in comparison to the LLL thus in disagreement with CF predictions. In conflict with CF predictions are also recent observations of FQHE at even denominator fractions in bilayer graphene including the most pronounced one at $\nu=-\frac{1}{2}$ (in the LLL for holes in the valence band) \cite{bil}. Though in graphene one deals with so-called relativistic version of the Landau level structure \cite{geb}, the FQHE in graphene is convergent with that one in the conventional semiconductor 2DEG. In particular, the incompressible FQHE state at $\nu=\frac{1}{2}$ has been observed formerly in conventional semiconductor bilayer Hall system \cite{2degbil,2degbill}. This state is distinct than the compressible Hall metal state observed in the monolayer system at $\nu=\frac{1}{2}$. The latter found an explanation in terms of CFs but the former one (in the bilayer system) not. 

All these listed above features indicate that the correlations in FQHE are distinct at various filling rates and still need clarification beyond the phenomenological CF model. Noticeably, none of the hypothetical secondary CF (with strong residual interaction) hierarchy states \cite{jain2014} is found in graphene as of yet, probably due to their lower stability in comparison to states from the main CF line as is also visible in experiments in the conventional 2DEG \cite{pan2003,5/13,4/11niezly}. 

The aim of the present paper is to summarize an alternative insight into correlated 2D Hall states which can be gained by utilization of the path integral quantization with topological methods \cite{wu,lwitt} taking advantage of nonlocal technique of braid group approach \cite{birman,jac-ws} allowing for exploration of odd topology of 2D space. It is commonly approved that behind the FQHE stands topological oddness of planar correlations of interacting electrons in 2D charged multiparticle system at magnetic field. The FQHE correlations are nonlocally conditioned by the topological factors and cannot be reduced to only local effects of interaction. 

In which manner the perpendicular magnetic field causes in a planar charged interacting multiparticle system the strongly nonlocal topological effect? The answer resolves itself to braid group analysis \cite{wu, lwitt, jac-ws} demonstrating constraints imposed on the {\it braid trajectories needed for the path integral quantization} by the cyclotron effect \cite{jac1}. The braids are homotopy classes of closed multiparticle trajectories (including particle indistinguishability) describing particle exchanges in the multiparticle configuration space. The configuration space and braid trajectories is not referred to real dynamics of quantum particles (without trajectories at all). Braids are classical objects used in topology terms upon the precisely defined braid group construction adjusted to requirements of the path integral quantization \cite{wu,imbo,birman,sud,jac-ws}. The braid group approach is not a single particle model, it is an essentially multiparticle nonlocal theory especially dedicated to recognize topological interparticle correlations. One-dimensional unitary representations (1DURs) of braids define weights for Feynman path integral of nonohomotopic path sectors (enumerated just by the braid group elements). Various 1DURs of the braid group allow for distinction of various quantum statistics of the same classical particles: fermions, bosons, anyons, composite fermions, composite bosons and composite anyons \cite{wu,sud,imbo,jac-ws}. As the multiparticle configuration spaces are not simply-connected then their homotopy groups (the $\pi_1$ homotopy groups for multiparticle configuration space are called braid groups \cite{jac-ws}) are nontrivial with also multiple 1DURs, especially rich in the case of 2D charged particles at magnetic field presence. The strong magnetic field discriminates particle trajectories in 2D and allows for braid particle exchanges exclusively when planar cyclotron orbits fit to particle separation \cite{jac1,jac2}. The resulting commensurability condition reproduces hierarchy of FQHE in full agreement with experimental observations. Related details are presented in the following paragraphs. 

\section{Commensurability condition and FQHE hierarchy in the LLL}

For the completely filled LLL, $\nu=\frac{N}{N_0}=1$, one deals with the commensurability condition:
\begin{equation}
\label{lll}
\frac{S}{N}= \frac{S}{N_0}=\frac{hc}{eB_0},
\end{equation}
where $S$ is the surface of the 2D sample, $B_0$ is the magnetic field corresponding to the completely filled LLL in the planar system with $N=N_0$ electrons, $N_0= \frac{B_0 S}{hc/e}$ is the degeneracy of LLs (for $B=B_O$) and $\frac{hc}{e}$ is the quantum of the magnetic field flux. The bare Landau kinetic energy $E_n=\hbar\omega_c(n+\frac{1}{2})$ with cyclotron energy $\omega_c=\frac{eB}{mc}$, here for $B=B_0$. The Eq. (\ref{lll}) states that the surface per single particle, $\frac{S}{N}$, equals to the {\it cyclotron orbit} size being the orbit size corresponding to single flux quantum, $\frac{hc}{eB_0}$. Despite the cyclotron orbits are meaningless quantumly, the surface of cyclotron orbit $\frac{hc}{eB_0}$ is still well defined and when this orbit fits to the surface per single particle $\frac{S}{N}$ we deal with the commensurability as given by Eq. (\ref{lll}). 

\begin{figure}[ht]
\centering
\makebox[\linewidth]{
\resizebox{0.8\textwidth}{!}{\includegraphics{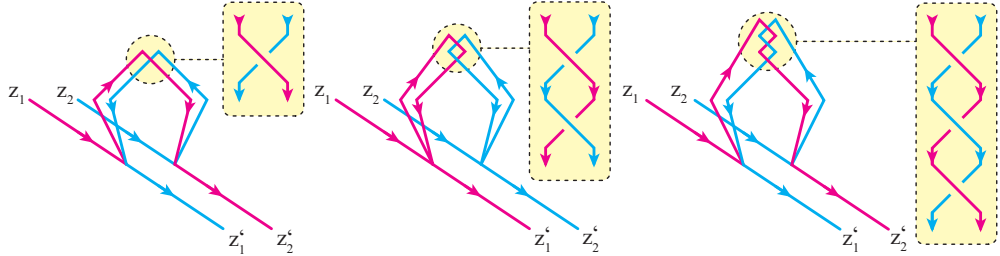}}
}
\caption{\label{fig25} Example of nonhomotopic trajectories obtained by addition of various braids to 2-particle trajectory}
\end{figure}

Above observations are especially important for the description of the multiparticle system in terms of the braid group. The so-called full braid group for $N$ particle system on the plane $R^2$ \cite{birman,jac-ws} defines exchanges between classical particles (which positions can be associated, on the other hand, with arguments of a multiparticle wave-function---any choice of its argument, $z_1,\dots,z_N$, corresponds to certain distribution of $N$ classical particles on the plane). The generators of the full braid group, $\sigma_i, \;i=1,\dots, N-1$, describe exchanges of neighboring particles ($i$th with $(i+1)$th at certain particle enumeration, arbitrary, however, due to particle indistinguishability imposed on the system by dividing of the configuration space of distinguishable particles by the permutation group $S_N$ \cite{birman,jac-ws}). 1DURs of the braid group (defined on the braid group generators $\sigma_i$) determine quantum particles (bosons, fermions, anyons, composite fermions, composite bosons and composite anyons \cite{jac-ws}). Various 1DURs, $e^{i\alpha_l}$, $\alpha_l \in[-\pi,\pi)$, $l$ enumerates braids, define different unitary weights for nonhomotopic (not continuously linked) sectors of trajectories in the domain of the Feynman path integral \cite{lwitt,wu},

\begin{equation}
\begin{array}{l}
I(z_1,\dots, z_N, t; z'_1,\dots, z'_N, t')\\
=\sum_{l\in \pi_1(\Omega)}e^{i\alpha_l}\int d\lambda_l e^{iS[\lambda_l(z_1,\dots, z_N, t; z'_1,\dots, z'_N, t')]/\hbar},\\
\end{array}
\end{equation}
where, $I(z_1,\dots, z_N, t; z'_1,\dots, z'_N, t')$ is the propagator, i.e., the matrix element of the evolution operator in position representation which determines the probability of quantum transition from the point $z_1,\dots, z_N$ in time instant $t$ to other point in the configuration space $z'_1,\dots, z'_N$ in time instant $t'$, $d\lambda_l$ is the measure in the path space sector enumerated by braid group element $l\in\pi_1(\Omega)$, $\pi_1(\Omega)$ is the first homotopy group of the configuration space $\Omega$ (it is just called the braid group), $\Omega= (M^N - \Delta)/S_N $, $M$ is 2D plane here, $M^N$ is $N$-fold normal product, $\Delta $ is the collection of diagonal points in the normal product (when at least two coordinates $z_i$ coincide) subtracted in order to assure particle number conservation, the quotient structure by the permutation group $S_N$ accounts for quantum indistinguishability of particles, $S[\lambda_l(z_1,\dots, z_N, t; z'_1,\dots, z'_N, t')]$ is the classical action for the trajectory $\lambda_l$ joining selected points in the configuration space $\Omega$ between time instances $t$, $t'$ and lying in $l$th sector of trajectory space. The whole space of trajectories is decomposed into disjoint sectors enumerated by braid group element index $l$ because to any trajectory at arbitrary time instant between $t$ and $t'$ one can adjoin the closed trajectory loop---the braid from the braid group. As braids are nonhomotopic thus resulted trajectories $\lambda_l$ are also nonhomotopic, i.e., they cannot be transformed one into another one in a continuous manner 
(in Fig. \ref{fig25} an example of nonhomotopic paths with additional braid loops are visualised for 2-particle system). This discontinuous decomposition of the domain of the path integral into disjoint sectors (topologically inequivalent) precludes a definition of the path measure $d\lambda$ uniformly on the whole space of paths and for each sector the measure $d\lambda_l$ must be defined separately and finally the contributions of all sectors must be summed with unitary factors $e^{i\alpha_l}$ (unitarity is caused by the causality constraint). It has been proved \cite{lwitt} that these unitary factors establish the 1DUR of the braid group. Distinct unitary weights in the path integral (i.e., distinct 1DURs of the braid group) determine different sorts of quantum particles corresponding to the same classical ones. Braids describe particle exchanges, thus their 1DURs assign quantum statistics. Equivalently, the 1DUR of a particular braid defines a phase shifts of the multiparticle wave function $\Psi(z_1,\dots ,z_N)$ when its arguments $z_1, \dots, z_N$ (classical coordinates of particles on the plane) mutually exchange according to this braid (let us emphasize that in 2D these exchanges are {\it not} permutations \cite{birman}). 

Turning back to the condition (\ref{lll}), we notice that it guarantees particle exchanges of uniformly distributed $N$ particles along cyclotron braid trajectories. At magnetic field presence no other trajectories are admitted, thus for the definition of the braid group the commensurability condition (\ref{lll}) must be fulfilled. The braid group can be then implemented and the statistics can be defined. The system is qauntumly correlated on whole plane and the correlation is expressed by the commensurability condition ($\ref{lll}$). 

 The condition (\ref{lll}) defines the correlation only when particles are uniformly distributed with fixed and constant separation. This is a case for triangle Wigner lattice of 2D electrons on the plane being the classical stable distribution for repulsing electrons located on uniform jellium at $T=0$. For free noninteracting particles the condition (\ref{lll}) does not define correlation because particle separation is not fixed in the gas and is not commensurate with the cyclotron orbit. The same happens when the temperature grows and the kinetic energy destroys the classical Wigner lattice---then the correlated IQHE is substituted by only completely filled LLL without any topological correlations. IQHE is the specific collective state for completely filled LLL. It is nonlocally correlated multiparticle state at low temperature when the kinetic energy is suppressed due the massive degeneracy of the LLL and the Wigner crystal distribution can be used as the classical model of the lowest energy state caused solely by interaction. When temperature reaches cyclotron energy $\hbar\omega_c$ this correlation would be disrupted. 

If $N_0$ is shifted (by the magnetic field change) then the correlation (\ref{lll}) is destroyed, because for $N\neq N_0$ the size of cyclotronic orbits does not fit to interparticle separation and electrons cannot exchange one with another one. Similar correlations for more complicated commensurability situations can be, however, recognized resulting in FQHE which will be described below. 

If the external magnetic field grows, $B>B_0$, then one deals with the fractional filling of the LLL $\nu<1$ ($N<N_0$ as the degeneracy has changed suitably to $B$, $N_0=\frac{BSe}{hc}$). One can determine the correlated states at fractional fillings generalizing the genuine pattern (\ref{lll}) assuring particle mutual exchanges despite the cyclotron orbits are too short to match neighboring particles at $B>B_0$. At fractional fillings of the LLL the cyclotron orbits $\frac{hc}{eB}$ are smaller than $\frac{S}{N}$ (as $B>B_0$) and cyclotron orbits cannot reach neighboring particles, which precludes ordinary cyclotron braid exchanges. For establishing of any correlated state the particle exchanges are, however, necessary to define statistics of quantum particles. This can be achieved by multiloop cyclotron orbits and related braid exchanges with additional loops. Exclusively in 2D multiloop cyclotron orbits have larger size in comparison to singlelooped ones at the same magnetic fields \cite{jac1,jac2}. This very peculiar property of a planar charged system follows from the distribution of the external field $B$ flux $\frac{BS}{N}$ per particle among all loops of the multiloop cyclotron orbit all located in the same plane (contrary to 3D, where each loop can lie on different surface and for 3D multiloop orbit the transpassing flux grows with number of loops, whereas in 2D does not). The condition for commensurability (\ref{lll}) at magnetic field $B> B_0$ attains thus the more general form:
\begin{equation}
\label{222}
\frac{BS}{N}=(q-1)\frac{hc}{ex}\pm\frac{hc}{ey},
\end{equation}
where:
$q$ is the number of loops of single cyclotron orbit, $q$ must be odd integer in order to ensure that the corresponding braid is a particle exchange. The braid generator with $n$ additional loops, $\sigma_i^n$, corresponds to $2n+1=q$ loop cyclotron orbit \cite{jac1,jac-ws}
 (at magnetic fields the braids in 2D are built form half-pieces of cyclotron orbits provided that these orbits accurately fit to neighboring particle separation at the uniform particle distribution). In the above formula 
$x\geq 1$ (integer) indicates the commensurability of $q-1$ single loops from $q$-loop cyclotron orbit to every $x$th particle on the plane (it follows from the relation $\frac{B'S}{N/x}=\frac{hc}{e}$, which is the condition (\ref{lll}) for fraction $N/x$ of particles). 
$y\geq x$ (also integer) indicates the commensurability of the last loop of the $q$-loop orbit with every $y$th particle. The sign 
$\pm$ indicates the same or opposite (of eight-figure-shape) orientation of the last i.e., $q$th loop.

\begin{figure*}[ht]
\centering
\makebox[\linewidth]{
\resizebox{1.0\textwidth}{!}{\includegraphics{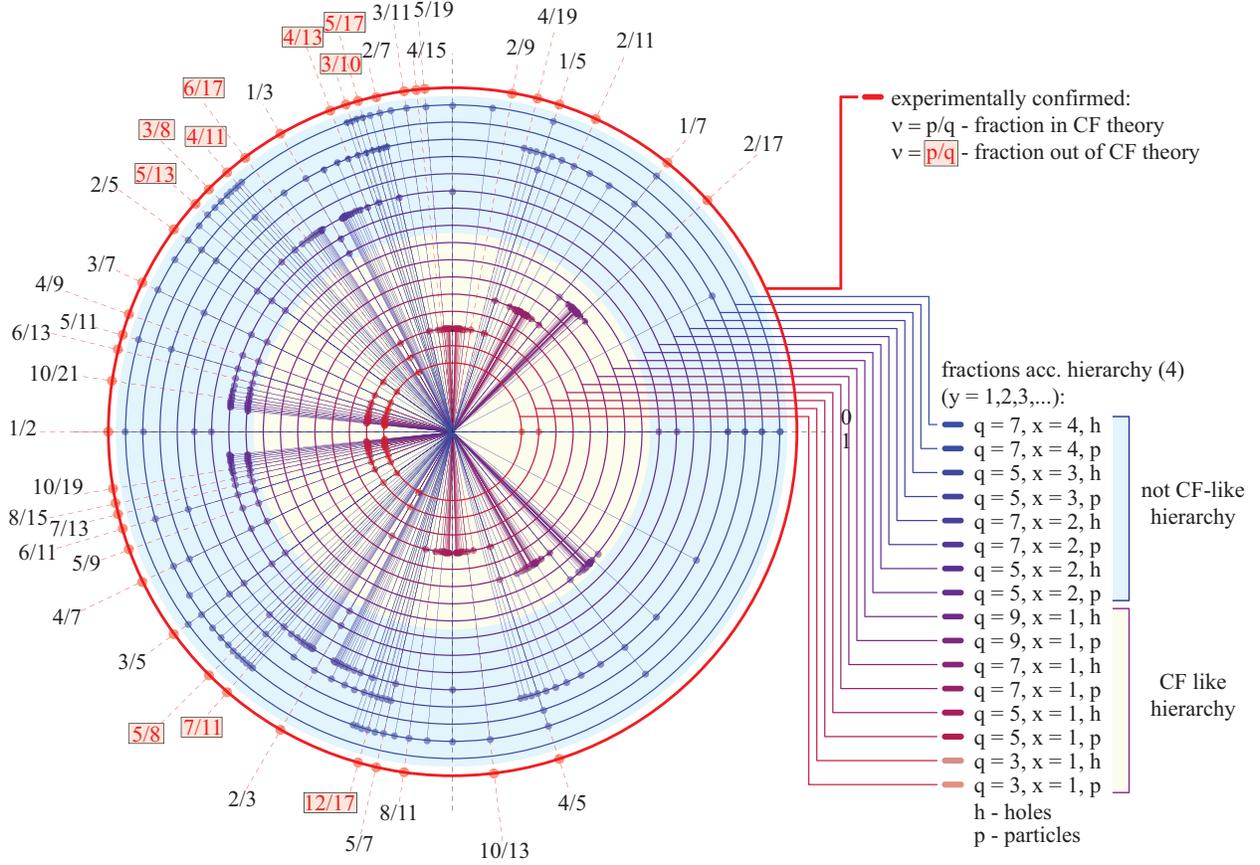}}
}
\caption{\label{fig1} Comparison of the hierarchy (\ref{444}) with all measured fractional filling rates for FQHE features in the LLL (spin polarized). The hierarchy series acc. (\ref{444}) for several $y$ each are displayed, filling rates beyond the main CF hierarchy are shown in red (Hall metal state fraction 1/2 is marked).}
\end{figure*}

In Eq. (\ref{222}) we note that its left-hand-side, $\frac{BS}{N}=\frac{N_0}{N}\frac{hc}{e}$, because $N_0=\frac{BS}{hc/e}$. Hence, from (\ref{222}) we obtain the following conditions for the general hierarchy of correlated states in the LLL describing the FQHE hierarchy, 
\begin{equation}
\label{444}
\begin{array}{l}
\nu=\frac{N}{N_0}=\frac{xy}{(q-1)y\pm x},\; \text{for band electrons},\\
\nu=1-\frac{xy}{(q-1)y\pm x},\; \text{for band holes},\\
\end{array} 
\end{equation}
For $x=1$ the hierarchy (\ref{444}) gives the FQHE hierarchy in the LLL derived by Jain \cite{jain} upon the effective model of CFs defined as electrons associated with localized on classical particles of auxiliary field flux tubes with even number of flux quanta each. For $x > 1$ the general hierarchy (\ref{444}) is beyond the ability of the CF model and gives ratios for FQHE observed in experiment outside the main CF hierarchy. The comparison with the experimental data is summarized in Fig. \ref{fig1}. 

It is clear thus that CFs simulate by fixed flux quanta additional loops for cyclotron orbits in the simplest case of the commensurability ($x=1$). Therefore CFs comprise nonlocal topological information on multiloop size in effective flux tubes imagined to be pinned to particles. The reduction of the external field by the average field of pinned to electrons flux tubes as assumed in CF model results in fact in enhancement of cyclotron orbit size what is actually needed for fulfilment of the braid commensurability requirement. The CF model is useful only in the simplest commensurability case ($x=1$) and breaks down in more complicated commensurability instances as given by Eq. (\ref{444}) for $x>1$. All experimentally observed FQHE features in the LLL displayed in Fig. \ref{fig2} may be linked to the newly derived hierarchy (\ref{444}). In red (in Fig. \ref{fig1}) are indicated filling rates out of main CF hierarchy, but they are visible in the experiment (cf. Fig. \ref{fig2}) and are successfully reproduced by the hierarchy (\ref{444}). Some examples of filling rates acc. to the hierarchy (\ref{444}) are collected in Tab. \ref{tab1}.

\begin{figure}[ht]
\centering
\resizebox{0.5\textwidth}{!}{\includegraphics{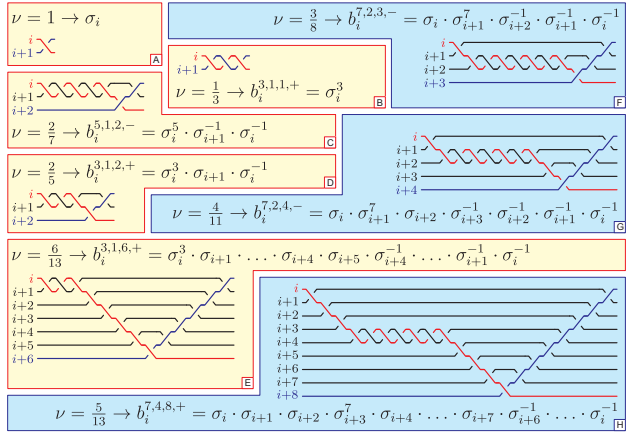}}
\caption{\label{fig3} The braid cyclotron subgroups generators for several selected filling fractions (blue background---examples of generators for filling fractions which cannot be derived using standard CF theory, yellow background---examples of fractions from CF-like hierarchy).}
\end{figure}

\begin{table}[th]
\centering
\begin{tabular}{|p{0.2cm}|p{0.2cm}|p{0.7cm}|p{5cm}|}
\hline
$q$&$x$&$y$& filling ratios acc. hierarchy (\ref{444}) \\
\hline
\scriptsize{3}& \scriptsize{1}& \scriptsize{1...10} & $\frac{1}{3}$, $\frac{2}{5}$, $\frac{3}{7}$, $\frac{4}{9}$, $\frac{5}{11}$, $\frac{6}{13}$, $\frac{7}{15}$, $\frac{8}{17}$, $\frac{9}{19}$, $\frac{10}{21}$, $\frac{2}{3}$, $\frac{3}{5}$, $\frac{4}{7}$, $\frac{5}{9}$, $\frac{6}{11}$, $\frac{7}{13}$, $\frac{8}{15}$, $\frac{9}{17}$, $\frac{10}{19}$, $\frac{11}{21}$, $...$\\
\hline
\scriptsize{5}& \scriptsize{1}& \scriptsize{1...5} & $\frac{1}{5}$, $\frac{2}{9}$, $\frac{3}{13}$, $\frac{1}{3}$, $\frac{2}{7}$, $\frac{3}{11}$, $\frac{4}{5}$, $\frac{7}{9}$, $\frac{10}{13}$, $\frac{2}{3}$, $\frac{5}{7}$, $\frac{8}{11}$, $...$\\
\hline
\scriptsize{5}& \scriptsize{2}& \scriptsize{2...5} & \textcolor{blue}{$\frac{2}{5}$}, \textcolor{blue}{$\frac{3}{7}$}, \textcolor{blue}{$\frac{4}{9}$}, \textcolor{blue}{$\frac{5}{11}$}, \textcolor{blue}{$\frac{2}{3}$}, \textcolor{blue}{$\frac{3}{5}$}, \textcolor{blue}{$\frac{4}{7}$}, \textcolor{blue}{$\frac{5}{9}$}, \textcolor{blue}{$\frac{6}{11}$}, \textcolor{blue}{$\frac{1}{3}$}, $...$\\
\hline
\scriptsize{5}& \scriptsize{3}& \scriptsize{3...5} & \textcolor{blue}{$\frac{3}{5}$}, \textcolor{red}{$\frac{12}{19}$}, \textcolor{red}{$\frac{15}{23}$}, \textcolor{blue}{$\frac{12}{13}$}, \textcolor{blue}{$\frac{15}{17}$}, \textcolor{blue}{$\frac{2}{5}$}, \textcolor{red}{$\frac{7}{19}$}, \textcolor{red}{$\frac{8}{23}$}, \textcolor{blue}{$\frac{1}{13}$}, \textcolor{red}{$\frac{2}{17}$}, $...$\\
\hline
\scriptsize{7}& \scriptsize{2}& \scriptsize{2...5} & \textcolor{blue}{$\frac{2}{7}$}, \textcolor{red}{$\frac{3}{10}$}, \textcolor{red}{$\frac{4}{13}$}, \textcolor{red}{$\frac{5}{16}$}, \textcolor{blue}{$\frac{2}{5}$}, \textcolor{red}{$\frac{3}{8}$}, \textcolor{red}{$\frac{4}{11}$}, \textcolor{red}{$\frac{5}{14}$}, \textcolor{blue}{$\frac{5}{7}$}, \textcolor{red}{$\frac{7}{10}$}, \textcolor{red}{$\frac{9}{13}$}, \textcolor{red}{$\frac{11}{16}$}, \textcolor{blue}{$\frac{3}{5}$}, \textcolor{red}{$\frac{5}{8}$}, \textcolor{red}{$\frac{7}{11}$}, \textcolor{red}{$\frac{9}{14}$}, $...$\\
\hline
\end{tabular}
\caption{Exemplary filling rates for FQHE in the LLL according to the hierarchy given by Eq. (\ref{444}) for various types of commensurability of cyclotron multilooped orbits; fractions obtained for $x>1$ (i.e., out of main line of CF model) are marked in red, though some of them coincide with CF-like hierarchy as an alternative commensurability (indicated in blue)}
\label{tab1} 
\end{table}

Note also that the limit $y\rightarrow \infty$ displays the hierarchy of the Hall metal exactly in the same manner as for the archetype of the Hall metal at $\nu=1/2$ (the last orbit is then infinite and fits to infinitely distant particles as in the normal Fermi liquid without any magnetic field \cite{st-no}). The general Hall metal hierarchy in the LLL has thus the form:
\begin{equation}
\label{555}
\begin{array}{l}
\nu= \frac{x}{q-1},\; \text{for electrons},\\
\nu=1-\frac{x}{q-1},\; \text{for holes}. \\
\end{array}
\end{equation}
Note that Hall metal correlation can manifest itself at fractions not necessarily with even denominators (for $x$ even, beyond the Jain CF concept), similarly as the hierarchy (\ref{444}) displays fractions both with odd and even denominators in compliance with the experimental observations (cf. Figs \ref{fig2} and \ref{fig1}). Some fractions are repeated in various lines of the general hierarchy (\ref{444}). This fact reveals the possibility of various types of commensurability of multiloop cyclotron orbits with interparticle separation $\frac{S}{N}$. The advantage of one commensurability over the others (alternative ones at the same filling ratio) is related with energy minimization, i.e., with the minimization of the Coulomb interaction being distinct for different commensurability patterns. 

\section{Wave functions for correlated FQHE states and Monte-Carlo Metropolis assessment of an activation energy}

For the simple line of the hierarchy (\ref{444}) with $x=y=1$, i.e., $\nu=\frac{1}{q}, \;q-odd$, the corresponding wave function has been given by Laughlin in the form \cite{laughlin2}:
\begin{equation}
\label{666}
\Psi_q(z_1,z_2,\dots, z_N)={\cal{A}}\prod\limits_{i,j,i>j}^{N,N}(z_i-z_j)^q e^{-\sum\limits_i^N\frac{|z_i|^2 }{4l^2}},
\end{equation}
where $z_i=x_i+iy_i$ is $i$th particle classical position (arguments of the quantum multiparticle wave function) on the complex plane, $l=\sqrt{\frac{hc}{eB}}$ is the magnetic length, and the product $\prod\limits_{i,j,i>j}^{N,N}(z_i-z_j)^q$ is the Jastrow polynomial---the generalization of the Vandermonde polynomial at $q=1$, {$\cal{A}$} is an appropriate normalization constant. The $q$-fold zero structure of the Jastrow polynomial keeps particles apart stronger than for Vandermonde polynomial, and thus diminishes the Coulomb interaction energy.

The function (\ref{666}) transforms itself according to the selected 1DUR of the cyclotron braid generated by $q$--loop cyclotron braid subgroup with generators $\sigma_i^q$. For the 1DUR of the full braid group given by $\sigma_i\rightarrow e^{i\alpha}$ with $\alpha =\pi$ (fermions) one can get the phase shift for the exchange of neighbors $i$th and $j=(i+1)$th as $(z_i-z_j)^q\rightarrow (z_j-z_i)^qe^{iq\pi}$ as required by the 1DUR of the generator $\sigma_i^q\rightarrow e^{iq\pi}$ (note that for more distant particles $i$th and $j$th, their exchange braid is given by the braid group element $(\sigma_i\cdot\dots\cdot\sigma_{j-2}\cdot\sigma_{j-1}\cdot\sigma_{j-2}^{-1}\cdot\dots\cdot\sigma_i^{-1})^q$ with 1DUR representation again equal to $e^{iq\pi}$ consistent with the form of the Laughlin function). Let us emphasize also, that the braid group generated by $\sigma_i^q$, $i=1,\dots,N-1$ is the subgroup of the full braid group generated by $\sigma_i$, $i=1,\dots,N-1$ (these subgroup is called as the cyclotron braid subgroup \cite{jac-ws}). The given above 1DUR of the cyclotron braid subgroup is the 1DUR of the full braid group, $\sigma_i\rightarrow e^{i\alpha}$, reduced to the subgroup. 

For the hierarchy (\ref{444}) the generators (describing elementary exchanges) of the appropriate more complicated cyclotron braid subgroups are defined as follows (for $\pm $ in (\ref{444})):
\begin{equation}
\label{777}
\begin{array}{l}
b_{i}^{q,x,y, +}=\\
(\! \sigma_i\! \cdot\! \sigma_{i+\!1}\! \cdot\! ...\! \cdot\! \sigma_{i+x-2}\! \cdot\! \sigma_{i+x-\!1}
\! \cdot\! \sigma_{i+x-2}^{-\!1}\! \cdot\! ...\! \cdot\! \sigma_{i+\!1}^{-\!1}\! \cdot\! \sigma_{i}^{-\!1})^{q-\!1} \\ \cdot\sigma_i\! \cdot\! \sigma_{i+\!1}\! \cdot\! ...\! \cdot\! \sigma_{i+y-2}\! \cdot\! \sigma_{i+y-\!1}\! \cdot\! \sigma_{i+y-2}^{-\!1}\! \cdot\! ...\! \cdot\! \sigma_{i+\!1}^{-\!1}\! \cdot\! \sigma_{i}^{-\!1},\\
\text{and}\\
b_{i}^{q,x,y, -}=\\
(\! \sigma_i\! \cdot\! \sigma_{i+\!1}\! \cdot\! ...\! \cdot\! \sigma_{i+x-2}\! \cdot\! \sigma_{i+x-\!1}\! \cdot\! \sigma_{i+x-2}^{-\!1}\! \cdot\! ...\! \cdot\! \sigma_{i+\!1}^{-\!1}\! \cdot\! \sigma_{i}^{-\!1})^{q-\!1} \\
\cdot (\! \sigma_i\! \cdot\! \sigma_{i+\!1}\! \cdot\! ...\! \cdot\! \sigma_{i+y-2}\! \cdot\! \sigma_{i+y-\!1}
\! \cdot\! \sigma_{i+y-2}^{-\!1}\! \cdot\! ...\! \cdot\! \sigma_{i+\!1}^{-\!1}\! \cdot\! \sigma_{i}^{-\!1})^{-\!1},\\
\end{array}
\end{equation}
with 1DURs (for $\alpha =\pi$) $ e^{iq\pi}$ (for $+$) and $e^{i(q-2)\pi}$ (for $-)$ (with supplement of the above notation for $x(y)=1$, 
$\sigma_i\cdot\sigma_{i+1}\cdot\dots \cdot\sigma_{i+x-2}\cdot \sigma_{i+x-1}
\cdot \sigma_{i+x-2}^{-1}\cdot \dots \cdot\sigma_{i+1}^{-1}\cdot\sigma_{i}^{-1}=\sigma_i$).
Examples of these braid generators are depicted in Fig. \ref{fig3}. 

The related modification of the Jastrow polynomial in the Laughlin function (\ref{666}) is thus as follows:
\begin{equation}
\label{888}
\begin{array}{l}
\Psi^{x,y,+}_q(z_1,z_2,\dots, z_N)=\\
\;\;\;{\cal{A}}\prod\limits_{\substack{\scriptscriptstyle{i,j=1;i<i\;\text{mod}\;x + (j-1)x}}}^{N,N/x}(z_i-z_{\scriptscriptstyle{i\;\text{mod}\;x + (j-1)x}})^{q-1}\\
\;\;\;\;\times\prod\limits_{\substack{\scriptscriptstyle{i,j=1;i<i\;\text{mod}\;y + (j-1)y}}}^{N,N/y}(z_i-z_{\scriptscriptstyle{i\;\text{mod}\;y + (j-1)y}}) e^{-\sum\limits_i^N\frac{|z_i|^2 }{4l^2}},\\
\Psi^{x,y,-}_q(z_1,z_2,\dots, z_N)=\\
\;\;\; {\cal{A}} \prod\limits_{\substack{\scriptscriptstyle{i,j=1;i<i\;\text{mod}\;x + (j-1)x}}}^{N,N/x}(z_i-z_{\scriptscriptstyle{i\;\text{mod}\;x + (j-1)x}})^{q-1}\\
\;\;\;\;\times\prod\limits_{\substack{\scriptscriptstyle{i,j=1;i<i\;\text{mod}\;y + (j-1)y}}}^{N,N/y}(z_{\scriptscriptstyle{i\;\text{mod}\;y + (j-1)y}} - z_i) e^{-\sum\limits_i^N\frac{|z_i|^2 }{4l^2}}.\\
\end{array}
\end{equation} 
The above functions for the Jain-like hierarchy ($x=1$) attains the form,
\begin{equation}
\label{999}
\begin{array}{l}
\Psi^{x=1,y,+}_q(z_1,z_2,\dots, z_N)=\\
\;\;\; {\cal{A}} \prod\limits_{\scriptscriptstyle{i,j=1,i<j}}^{N,N}(z_i-z_j)^{q-1}\\
\;\;\;\;\times\prod\limits_{\substack{\scriptscriptstyle{i,j=1;i<i\;\text{mod}\;y + (j-1)y}}}^{N,N/y}(z_i-z_{\scriptscriptstyle{i\;\text{mod}\;y + (j-1)y}}) e^{-\sum\limits_i^N\frac{|z_i|^2 }{4l^2}},\\
\Psi^{x=1,y,-}_q(z_1,z_2,\dots, z_N)=\\
\;\;\; {\cal{A}} \prod\limits_{\scriptscriptstyle{i,j=1,i<j}}^{N,N}(z_i-z_j)^{q-1}\\
\;\;\;\;\times\prod\limits_{\substack{\scriptscriptstyle{i,j=1;i<i\;\text{mod}\;y + (j-1)y}}}^{N,N/y}(z_{\scriptscriptstyle{i\;\text{mod}\;y + (j-1)y}} - z_i) e^{-\sum\limits_i^N\frac{|z_i|^2 }{4l^2}}.\\
\end{array}
\end{equation} 

The functions (\ref{888}) are proposed as the trial wave functions for correlated states for filling rates (\ref{444}) for which elementary exchanges of particles are defined by braids (\ref{777}) and generalize the Laughlin function (\ref{666}) for the case when $x,y>1$. 

The energy gain in the Laughlin state is due to the reducing of the Coulomb repulsion energy $\sum\limits_{i.j,i>j}^{N,N}\frac{e^2}{|z_i-z_j|}$. It is clear that the energy reducing with the function 
(\ref{888}) is the weaker the higher $x$ is (for the same $q$ and $y$). It follows from the dilution of correlated particles for $x>1$ (the correlation concerns every $x$th electron only) as expressed in modified Laughlin-type function (\ref{888}) by reducing of the domain of the product. This leads to the diminishing of the repulsion energy gain due to the averaging of the Coulomb energy, $\sum\limits_{i.j,i>j}^{N,N}\frac{e^2}{|z_i-z_j|}$, with the wave function (\ref{888}) instead of (\ref{999}) (or (\ref{666})) because $q-1$ fold zero in these functions prevents approaching not all electrons in the case of function (\ref{888}) but only its $1/x$ fraction (opposite to the case of function (\ref{666}) or (\ref{999}) for which $x=1$). Therefore more stable are states with lower $x$. Thus states with $x=1$ energetically prevail over states with $ x>1$ and are more stable.

 To confront the activation energy values obtained from numerical exact diagonalization of electron interaction for different fillings corresponding to FQHE in small models \cite{jain-exact-2015}, the numerical calculation of expectation value of interaction energy for newly proposed functions (\ref{888}) and (\ref{999}) were performed by the Monte Carlo integration method upon the Metropolis scheme \cite{montecarlo1,montecarlo2,metropolis}.
The Monte Carlo Metropolis method \cite{metropolis} is especially effective in assessment of the multi-argument integrals (we have utilized this method to models with 200 particles in circular symmetry of boundary conditions). For 400-fold argument (for 200 particles on 2D plane) of the multiple integral the standard Monte Carlo procedure are ineffective but the required accuracy can be achieved by utilization of appropriately modified version originally formulated by Metropolis for large scale chemical thermodynamic modeling \cite{metropolis}. To quantum Hall systems the method has been adopted in \cite{montecarlo1} but ranged to only analysis of Laughlin type states for $\nu=\frac{1}{3}$ and $\frac{1}{5}$. We have developed the method of Metropolis also to other states from FQHE hierarchy and related to trial wave functions given by Eqs. (\ref{888}) and (\ref{999}).
Effectiveness of the applied multi-variable integration method consists in searching of local maxima of the integrand expression by the self-organized optimal random walk. As the integrand is proportional to the density of probability of distribution of $N$ particles, the resulted pattern of local maxima distribution (probability distribution for particle positions) reveals correlations imposed on the multiparticle system by the wave function under examination. These patterns of correlations can be easy compared with respect to the expectation value of particle electric interaction. The final averaged value of Coulomb energy, specific for each trial wave function form, is obtained by averaging over $10^7$ repetitions of the maxima searching steps avoiding few thousands initial steps to suppress an initial configuration influence. Next the procedure to assess of the expectation value of interaction energy in the thermodynamic limit is performed repeating the calculus for different $N$ and by the linear extrapolation with respect to $\frac{1}{\sqrt{N}}\rightarrow 0$ \cite{montecarlo1} in the range for $N=4$ up to $200$ in our study. 
Some exemplary results revealing the very good consistence of the determined Monte Carlo Metropolis expectation value of interparticle interaction energy (including interaction with positive jellium) with the energy found by the exact diagonalization of the Coulomb interaction in small models for FQHE \cite{jain-exact-2015} are presented in Table \ref{tab2}.

\begin{table}[th]
\centering
\begin{tabular}{|p{0.2cm}|p{0.2cm}|p{0.2cm}|p{2.5cm}|p{1.7cm}|p{1.7cm}|}
\hline
$q$ & $x$ & $y$ & \scriptsize{hierarchy fraction, $\nu = N/N_0$} & \scriptsize{energy from Monte Carlo simulation for functions, according to Eq. (\ref{888},\ref{999})} & \scriptsize{energy from exact diagonalization \cite{jain-exact-2015}}\\
\hline
\scriptsize{3}&\scriptsize{1}&\scriptsize{2}&$\frac{2\cdot 1}{(3-1)\cdot 2+1}=\frac{2}{5}$&\scriptsize{$-0.432677$}&\scriptsize{$-0,432804$}\\
\hline
\scriptsize{3}&\scriptsize{1}&\scriptsize{3}&$\frac{3\cdot 1}{(3-1)\cdot 3+1}=\frac{3}{7}$&\scriptsize{$-0.441974$}&\scriptsize{$-0,442281$}\\
\hline
\scriptsize{3}&\scriptsize{1}&\scriptsize{4}&$\frac{4\cdot 1}{(3-1)\cdot 4+1}=\frac{4}{9}$&\scriptsize{$-0.446474$}&\scriptsize{$-0,447442$}\\
\hline
\scriptsize{3}&\scriptsize{1}&\scriptsize{5}&$\frac{5\cdot 1}{(3-1)\cdot 5+1}=\frac{5}{11}$&\scriptsize{$-0.451056$}&\scriptsize{$-0,450797$}\\
\hline
\scriptsize{5}&\scriptsize{1}&\scriptsize{2}&$\frac{2\cdot 1}{(5-1)\cdot 2+1}=\frac{2}{9}$&\scriptsize{$-0.342379$}&\scriptsize{$-0,342742$}\\
\hline
\scriptsize{5}&\scriptsize{1}&\scriptsize{3}&$\frac{3\cdot 1}{(5-1)\cdot 3+1}=\frac{3}{13}$&\scriptsize{$-0.348134$}&\scriptsize{$-0,348349$}\\
\hline
\scriptsize{5}&\scriptsize{1}&\scriptsize{4}&$\frac{4\cdot 1}{(5-1)\cdot 4+1}=\frac{4}{17}$&\scriptsize{$-0.351857$}&\scriptsize{$-0,351189$}\\
\hline
\end{tabular}
\caption{Comparison of energy values obtained by exact diagonalization and by Monte Carlo simulation for some exemplary filling fractions for FQHE (Monte Carlo Metropolis simulation for trial wave functions derived by the cyclotron braid group method given by Eqs. (\ref{888}) and (\ref{999}), for $N= 200$ particles).}
\label{tab2} 
\end{table}

Turning back to the commensurability condition (\ref{444}) it should be commented that for multiloop cyclotron orbits, none of the loop cannot be featured in general, thus each loop may be accommodated to the particle separation independently. Thus, for $q$-looped orbit one would deal with the ordered series $x_1\leq x_2\leq \dots \leq x_q$ simplified in (\ref{444}) to $x_1=\dots=x_{q-1}=x, \; x_q=y$. Apparently, the Coulomb repulsion minimization prefers $x_1=\dots=x_{q-1}=x$ for which the minimization domain restriction (resulting in weaker interaction energy reducing) is more convenient than for distinct distributions of $x_i$. This explains the choice of the uniform behavior od $q-1$ loops (i.e., $x_1=\dots =x_{q-1}=x$) but this is not a rule and for many fractions various energetically competitive commensurability opportunities might be considered in principle.

The another observation related to various types of correlation identified by the commensurability criterion agrees with experimental data for the longitudinal resistivity $R_{xx}$ (cf. Fig. \ref{fig2}) which is zero for states with all correlated particles (i.e., with $x=1$), whereas the residual its value grows with $x>1$ probably due to scattering on portion of non-correlated electrons. 

\section{Symmetry of wave functions for FQHE states and violation of fermionic antisymmetry at some filling fractions}

1DURs of cyclotron braid subgroup generated by the commensurability 
condition (\ref{222}) defines symmetry of the corresponding wave function for multiparticle correlated state referred to FQHE at particular filling rates (\ref{444}) given by the commensurability constraint. The wave function must transform according to 1DUR of the braid when the argument of the wave function $z_1,\dots, z_N$ change position on the plane according to this particular braid. This symmetry property together with condition that the wave function in the LLL must be a holomorphic function (then uniquely defined by its nodes), the shape of the function can be determined up to an invariant to braid transformation term. This term must be the same one as for the gas system because the gas wave functions span the Hilbert space for wave functions with interactions. This invariant term has the form of exponent, $exp(-\sum_{i=1}^N |z_i|^2/4l^2)$. Thus, all the symmetry resolves itself to the polynomial part of the wave function as have been defined for various commensurabilities (\ref{222}) by the expressions (\ref{888}) or (\ref{999}). 

It must be emphasized that the arbitrariness in choice of 1DURs corresponding to a plethora of composite anyons (being at the presence of strong magnetic the generalization of ordinary anyons in planar system without the quantizing magnetic field) can be reduced by assumption of fermion type particles corresponding to original covering full braid group. In this way we may address only to composite fermions and their wave function symmetry (the name related to CFs may be justified by the limit $x=1$ in the commensurability condition (\ref{222})), i.e., 1DUR$= e^{i\pi}$
for original generators $\sigma_i$ of the full braid group. Though the resulted 1DURs for cyclotron subgroups have the form $e^{ip\pi}$ with $p$ odd integer, the corresponding wave functions have not fermionic antisymmetry property in general. This is caused by the fact that new generators of corresponding cyclotron subgroups do not define simple exchange of function variables as it was in 3D space described by the permutation group. In 2D braid groups and their cyclotron subgroups are far more complicated than the permutation group. Therefore the wave function satisfying symmetry requirements imposed by the 1DUR of the cyclotron subgroup induced by the particular commensurability condition has not a simple fermionic antisymmetry property (as visible in expressions (\ref{888}) and (\ref{999})). This feature is first time clearly explained in terms of presented above formalism but contains a significant exceptional conclusion on violation of fermionic antisymmetry property of electrons creating many FQHE states. This is a fundamental difference with respect to 3D correlated states described obligatory by the antisymmetric functions for fermions. In 2D we have proved that this antisymmetry can be violated. 

It must be emphasized that we have thus elucidated an embarrassing fact upon the CF theory when the conjecture on the form of the corresponding wave function has been formulated by utilization of wave functions for completely filled higher LLs \cite{jain2007}. To avoid poles present in wave functions for LLs, some heuristic procedures were then applied to model finally the holomorphic multiparticle function for fractional filling of the LLL without any poles. Despite many attempts none rigorous method for this artificial procedure of projection onto LLL has been proved and the method was uncertain and frequently searched empirically by energy minimization. The procedure of this projection onto LLL violates, however in an uncontrolled manner, the fermionic antisymmetry of electrons described by the final function. Nevertheless, the rigorous definition of the wave function symmetry according to the appropriate 1DUR of the cyclotron braid subgroup generated by a specific commensurability condition solves the problem of symmetry of electrons at FQHE accurately and completely. Worth noting is the fact that it is a sole case of violation of fermionic antisymmetry of electrons in condensed matter. This significant and fundamental quantum property is born by also exceptional multiloop cyclotron topology and related nonlocal correlations exclusively in 2D space. 

\section{Summary} 

In conclusion we state that the commensurability condition selects the filling ratios for correlated states which coincide very well with the experimentally observed FQHE hierarchy. All experimentally noted fractional fillings for incompressible states in the LLL are predicted theoretically in this way. Worth noting is that all experimentally observed fractions for FQHE beyond the so-called main CF hierarchy are reproduced by the braid group commensurability condition (in Fig. \ref{fig1} twelve such fractions are indicated). Remarkably, all the experimentally observed filling fractions for FQHE and for Hall metal states are accounted for in the hierarchy given by Eqs. (\ref{444}) and (\ref{555}) in the LLL. The presented braid group topological approach to FQHE hierarchy is complete, i.e., the cyclotron braid subgroup appropriate to each selected fraction is defined by the explicit construction of relevant subgroup generators in compliance with the specific commensurability criterion. The forms of generators allow then for shaping of the generalization of Laughlin function polynomial multiplier which must transform according to the 1DUR of related cyclotron braid generators. Utilizing the form of this wave function generalization one can conclude on comparative stability of various correlated states via assessment of their energies, i.e., of interaction energy averaged over theses trial wave functions. The averaged energies obtained by Monte-Carlo Metropolis method of multivariable function integration are in agreement with experimental data and coincide with exact diagonalization of Coulomb energy in small models. 

Let us emphasize that the commensurability condition based on the braid group approach has nonlocal and topological character. The role of the Coulomb interaction is of the primary importance for this method. FQHE is the nonperturbative effect induced by the Coulomb interaction of electrons in planar geometry at strong perpendicular magnetic field presence. In this context worth mentioning is that in Hall systems the Coulomb interaction does not lead to a continuous mass operator which precludes the quasiparticle concept \cite{landau1972,abrikosov1975}. Instead we deal here with particle separation quantization \cite{laughlin1,laughlin2} expressed in the form of the famous Laughlin function and Haldane interaction pseudopotentials \cite{pinczuk1997,hh1}. The related nonlocal and nonperturbative correlations cannot be expressed by any local or single-particle approach. The conventional CF model \cite{jain2007} though uses single-particle notion of effective composite particles is in fact also nonlocal one due to involvement of effective auxiliary field flux quanta pinned to electrons. This construction is essentially nonlocal and nonperturbative and displays the topological deep property of the 2D charged system in magnetic field clearly different than the concept of quasiparticles common in solids \cite{abrikosov1975}. CFs are neither quasiparticles nor effective single-particle excitations in the Hall system. They reveal rather the long range correlation between 2D electrons expressed by auxiliary condition of quantization of fictitious flux pinned to each electron in order to form the composite particle. The character of this long range correlation is clearly explained in terms of the braid group approach and related commensurability condition allowing for clarifying of the structure of CFs and of constraints imposed on the CF model, on the other hand. This is especially well visible in the manifestation of FQHE at these fractions which cannot be explained by the conventional CF model but are quite naturally embraced by the commensurability condition (e.g., LLL fractions with even denominator like $\frac{3}{8}$, $\frac{3}{10}$, or LLL fraction $\frac{1}{2}$ in bilayer Hall system revealing FQHE and not Hall metal as in monolayer system \cite{bil,jacak-royal}). The braid group commensurability condition allows for explanation of full experimentally observed hierarchy of filling fractions corresponding to FQHE including those admitted by the CF model as a particular limiting situation. The convincing example is also the possibility of an explanation in braid group terms of the unconventional FQHE observed in the LLL of bilayer graphene at even denominator fractions \cite{bil} beyond the prediction of the CF model \cite{jacak-royal}. Note that the presented commensurability braid group approach is efficient also in higher LLs \cite{d6,d7} where constraints limiting applicability of the CF model are even more severe than those in the LLL \cite{jac-zhetf-hll}.

The most important feature related with symmetry of carriers involved in correlations of FQHE and revealed by the developed approach of cyclotron braid subgroups induced by the specific commensurability condition in 2D charged interacting systems at magnetic field presence is a possible violation of fermionic antisymmetry property of 2D electrons. This fundamental observation is unique in condensed matter and closely related with specific topology of 2D space resulted in FQHE. 

\acknowledgments
Supported by the NCN projects P.2011/02/A/ST3/00116 and P.2016/21/D/ST3/00958.


\begin{thebibliography}{10}

\bibitem{tsui1982}
D.~C. Tsui, H.~L. St{\"o}rmer, and A.~C. Gossard.
\newblock Two-dimensional magnetotransport in the extreme quantum limit.
\newblock {\em Phys. Rev. Lett.}, 48:1559, 1982.

\bibitem{laughlin2}
R.~B. Laughlin.
\newblock Anomalous quantum {H}all effect: an incompressible quantum fluid with
  fractionally charged excitations.
\newblock {\em Phys. Rev. Lett.}, 50:1395, 1983.

\bibitem{laughlin1}
R.~B. Laughlin.
\newblock Quantized motion of three two-dimensional electrons in a strong
  magnetic field.
\newblock {\em Phys. Rev. B}, 27:3383, 1983.

\bibitem{pan2003}
W.~Pan, H.~L. St{\"o}rmer, D.~C. Tsui, L.~N. Pfeiffer, K.~W. Baldwin, and K.~W.
  West.
\newblock Fractional quantum {H}all effect of composite fermions.
\newblock {\em Phys. Rev. Lett.}, 90:016801, 2003.

\bibitem{jain-n}
J.~K. Jain.
\newblock A note contrasting two microscopic theories of the fractional quantum
  hall effect.
\newblock {\em Indian Journal of Physics}, 88:915, 2014.

\bibitem{lls}
J.~P. Eisenstein, M.~P. Lilly, K.~B. Cooper, L.~N. Pfeiffer, and K.~W. West.
\newblock New physics in high {L}andau levels.
\newblock {\em Physica E}, 6:29, 2000.

\bibitem{ll8/3-1}
M.~Dolev, Y.~Gross, R.~Sabo, I.~Gurman, M.~Heiblum, V.~Umansky, and D.~Mahalu.
\newblock Characterizing neutral modes of fractional states in the second
  {L}andau level.
\newblock {\em Phys. Rev. Lett.}, 107:036805, 2011.

\bibitem{5/2-1}
R.~L. Willett.
\newblock The quantum {H}all effect at 5/2 filling factor.
\newblock {\em Rep. Prog. Phys.}, 76:076501, 2013.

\bibitem{ll8/3-2}
W.~Pan, K.~W. Baldwin, K.~W. West, L.~N. Pfeiffer, and D.~C. Tsui.
\newblock Spin transition in the $\nu =8/3$ fractional quantum {H}all effect.
\newblock {\em Phys. Rev. Lett.}, 108:216804, 2012.

\bibitem{xia2004}
J.~S. Xia, W.~Pan, C.~L. Vicente, E.~D. Adams, N.~S. Sullivan, H.~L.
  St{\"o}rmer, D.~C. Tsui, L.~N. Pfeiffer, K.~W. Baldwin, and K.~W. West.
\newblock Electron correlation in the second {L}andau level: a competition
  between many nearly degenerate quantum phases.
\newblock {\em Phys. Rev. Lett.}, 93:176809, 2004.

\bibitem{amet}
F.~Amet, A.~J. Bestwick, J.~R. Williams, L.~Balicas, K.~Watanabe, T.~Taniguchi,
  and D.~{Goldhaber-Gordon}.
\newblock Composite fermions and broken symmetries in graphene.
\newblock {\em Nat. Commun.}, 6:5838, 2015.

\bibitem{dean2011}
C.~R. Dean, A.~F. Young, P.~Cadden-Zimansky, L.~Wang, H.~Ren, K.~Watanabe,
  T.~Taniguchi, P.~Kim, J.~Hone, and K.~L. Shepard.
\newblock Multicomponent fractional quantum {H}all effect in graphene.
\newblock {\em Nature Physics}, 7:693, 2011.

\bibitem{feldman}
B.~E. Feldman, B.~Krauss, J.~H. Smet, and A.~Yacoby.
\newblock Unconventional sequence of fractional quantum {H}all states in
  suspended graphene.
\newblock {\em Science}, 337:1196, 2012.

\bibitem{feldman2}
B.~E. Feldman, A.~J. Levin, B.~Krauss, D.~A. Abanin, B.~I. Halperin, J.~H.
  Smet, and A.~Yacoby.
\newblock Fractional quantum {H}all phase transitions and four-flux states in
  graphene.
\newblock {\em Phys. Rev. Lett.}, 111:076802, 2013.

\bibitem{bil}
D.~K. Ki, V.~I. Falko, D.~A. Abanin, and A.~Morpurgo.
\newblock Observation of even denominator fractional quantum {H}all effect in
  suspended bilayer graphene.
\newblock {\em Nano Lett.}, 14:2135, 2014.

\bibitem{kou2014science}
A.~Kou, B.~E. Feldman, A.~J. Levin, B.~I. Halperin, K.~Watanabe, T.~Taniguchi,
  and A.~Yacoby.
\newblock Electron-hole asymmetric integer and fractional quantum hall effect
  in bilayer graphene.
\newblock {\em Science}, 345:55, 2014.

\bibitem{maher2014science}
Patrick Maher, Lei Wang, Yuanda Gao, Carlos Forsythe, Takashi Taniguchi, Kenji
  Watanabe, Dmitry Abanin, Zlatko Papi{\'c}, Paul Cadden-Zimansky, James Hone,
  Philip Kim, and Cory~R. Dean.
\newblock Tunable fractional quantum hall phases in bilayer graphene.
\newblock {\em Science}, 345:61, 2014.

\bibitem{kim2015nanolett}
Youngwook Kim, Dong~Su Lee, Suyong Jung, Viera Skákalová, T.~Taniguchi,
  K.~Watanabe, Jun~Sung Kim, and Jurgen~H. Smet.
\newblock Fractional quantum hall states in bilayer graphene probed by
  transconductance fluctuations.
\newblock {\em Nano Lett.}, 15:7445, 2015.

\bibitem{diankov2016naturecomm}
Georgi Diankov, Chi-Te Liang, François Amet, Patrick Gallagher, Menyoung Lee,
  Andrew~J. Bestwick, Kevin Tharratt, William Coniglio, Jan Jaroszynski, Kenji
  Watanabe, Takashi Taniguchi, and David Goldhaber-Gordon.
\newblock Robust fractional quantum hall effect in the n=2 landau level in
  bilayer graphene.
\newblock {\em Nature Comm.}, 7:13908, 2016.

\bibitem{geb}
M.~O. Goerbig.
\newblock Electronic properties of graphene in a strong magnetic field.
\newblock {\em Rev. Mod. Phys.}, 83:1193, 2011.

\bibitem{jain}
J.~K. Jain.
\newblock Composite-fermion approach for the fractional quantum {H}all effect.
\newblock {\em Phys. Rev. Lett.}, 63:199, 1989.

\bibitem{jain2007}
J.~K. Jain.
\newblock {\em Composite Fermions}.
\newblock Cambridge UP, Cambridge, 2007.

\bibitem{jain2014}
S.~Mukherjee, S.~S. Mandal, {Y.H.} Wu, A.~W{\'o}js, and J.~K. Jain.
\newblock Enigmatic 4/11 state: A prototype for unconventional fractional
  quantum hall effect.
\newblock {\em Phys. Rev. Lett.}, 112:016801, 2014.

\bibitem{d1}
P.~Sitko, K.-S. Yi, and J.~J. Quinn.
\newblock Composite fermion hierarchy: Condensed states of composite fermion
  excitations.
\newblock {\em Phys. Rev. B}, 56:12417, 1997.

\bibitem{d2}
K.~Park and J.~K. Jain.
\newblock Mixed states of composite fermions carrying two and four vortices.
\newblock {\em Phys. Rev. B}, 62:R13274, 2000.

\bibitem{d3}
A.~W\'ojs, K.-S. Yi, and J.~J. Quinn.
\newblock Fractional quantum hall states of clustered composite fermions.
\newblock {\em Phys. Rev. B}, 69:205322, 2004.

\bibitem{d4}
W.~Pan, K.~W. Baldwin, K.~W. West, L.~N. Pfeiffer, and D.~C. Tsui.
\newblock Fractional quantum hall effect at landau level filling
  $\ensuremath{\nu}=4/11$.
\newblock {\em Phys. Rev. B}, 91:041301, 2015.

\bibitem{d5}
N.~Samkharadze, I.~Arnold, L.~N. Pfeiffer, K.~W. West, and G.~A. Cs\'athy.
\newblock Observation of incompressibility at $\ensuremath{\nu}=4/11$ and
  $\ensuremath{\nu}=5/13$.
\newblock {\em Phys. Rev. B}, 91:081109, 2015.

\bibitem{abrikosov1975}
A.~A. Abrikosov, L.~P. Gorkov, and I.~E. Dzialoshinskii.
\newblock {\em Methods of Quantum Field Theory in Statistical Physics}.
\newblock Dover Publ. Inc., Dover, 1975.

\bibitem{2degbil}
Y.~W. Suen, L.~W. Engel, M.~B. Santos, M.~Shayegan, and D.~C. Tsui.
\newblock Observation of a $\nu$=1/2 fractional quantum {H}all state in a
  double-layer electron system.
\newblock {\em Phys. Rev. Lett.}, 68:1379, 1992.

\bibitem{2degbill}
J.~P. Eisenstein, G.~S. Boebinger, L.~N. Pfeiffer, K.~W. West, and S.~He.
\newblock New fractional quantum {H}all state in double-layer two-dimensional
  electron systems.
\newblock {\em Phys. Rev. Lett.}, 68:1383, 1992.

\bibitem{5/13}
N.~Samkharadze, I.~Arnold, L.~N. Pfeiffer, K.~W. West, and G.~A. Cs\'athy.
\newblock Observation of incompressibility at $\ensuremath{\nu}=4/11$ and
  $\ensuremath{\nu}=5/13$.
\newblock {\em Phys. Rev. B}, 91:081109, 2015.

\bibitem{4/11niezly}
W.~Pan, K.~W. Baldwin, K.~W. West, L.~N. Pfeiffer, and D.~C. Tsui.
\newblock Fractional quantum hall effect at landau level filling 4/11.
\newblock {\em Phys. Rev. B}, 91:041301(R), 2015.

\bibitem{wu}
Y.~S. Wu.
\newblock General theory for quantum statistics in two dimensions.
\newblock {\em Phys. Rev. Lett.}, 52:2103, 1984.

\bibitem{lwitt}
M.~G. Laidlaw and C.~M. DeWitt.
\newblock Feynman functional integrals for systems of indistinguishable
  particles.
\newblock {\em Phys. Rev. D}, 3:1375, 1971.

\bibitem{birman}
J.~S. Birman.
\newblock {\em Braids, Links and Mapping Class Groups}.
\newblock Princeton UP, Princeton, 1974.

\bibitem{jac-ws}
J.~Jacak, R.~Gonczarek, L.~Jacak, and I.~J{\'o}{\'z}wiak.
\newblock {\em Application of Braid Groups in {2D} Hall System Physics:
  Composite Fermion Structure}.
\newblock World Scientific, 2012.

\bibitem{jac1}
J.~Jacak, I.~J\'o\'zwiak, and L.~Jacak.
\newblock New implementation of composite fermions in terms of subgroups of a
  braid group.
\newblock {\em Phys. Lett. A}, 374:346, 2009.

\bibitem{imbo}
T.~D. Imbo, C.~S. Imbo, and C.~S. Sudarshan.
\newblock Identical particles, exotic statistics and braid groups.
\newblock {\em Phys. Lett. B}, 234:103, 1990.

\bibitem{sud}
E.~C.~G. Sudarshan, T.~D. Imbo, and T.~R. Govindarajan.
\newblock Configuration space topology and quantum internal symmetries.
\newblock {\em Phys. Lett. B}, 213:471, 1988.

\bibitem{jac2}
J.~Jacak, I.~J\'o\'zwiak, L.~Jacak, and K.~Wieczorek.
\newblock Cyclotron braid group structure for composite fermions.
\newblock {\em J. Phys: Cond. Matt.}, 22:355602, 2010.

\bibitem{st-no}
H.~L. Stormer, R.~R. Du, W.~Kang, D.~C. Tsui, L.~N. Pfeiffer, K.~W. Baldwin,
  and K.~W. West.
\newblock The fractional quantum {H}all effect in a new light.
\newblock {\em Semicond. Sci. Technol.}, 9:1853, 1994.

\bibitem{jain-exact-2015}
Ajit~C. Balram, Csaba T\ifmmode~\mbox{\H{o}}\else \H{o}\fi{}ke, A.~W\'ojs, and
  J.~K. Jain.
\newblock Fractional quantum hall effect in graphene: Quantitative comparison
  between theory and experiment.
\newblock {\em Phys. Rev. B}, 92:075410, 2015.

\bibitem{montecarlo1}
Orion Ciftja and Carlos Wexler.
\newblock Monte carlo simulation method for laughlin-like states in a disk
  geometry.
\newblock {\em Phys. Rev. B}, 67:075304, 2003.

\bibitem{montecarlo2}
R.~Morf and B.~I. Halperin.
\newblock Monte carlo evaluation of trial wavefunctions for the fractional
  quantized hall effect: Spherical geometry.
\newblock {\em Z. Phys. B Condensed Matter}, 68:391, 1987.

\bibitem{metropolis}
N.~Metropolis, A.~W. Rosenbluth, M.~N. Rosenbluth, A.~M. Teller, and E.~Teller.
\newblock Equation of state calculations by fast computing machines.
\newblock {\em J. Chem. Phys.}, 21:1087, 1953.

\bibitem{landau1972}
L.~D. Landau and E.~M. Lifshitz.
\newblock {\em Quantum Mechanics: Non-relativistic Theory}.
\newblock Nauka, Moscow, 1972.

\bibitem{pinczuk1997}
S.~{Das Sarma} and A.~Pinczuk.
\newblock {\em Perspectives in Quantum Hall Effects: Novel Quantum Liquids in
  Low-Dimensional Semiconductor Structures}.
\newblock Wiley, New York, 1997.

\bibitem{hh1}
F.~D.~M. Haldane.
\newblock Fractional quantization of the {H}all effect:{A} hierarchy of
  incompressible quantum fluid states.
\newblock {\em Phys. Rev. Lett.}, 51:605, 1983.

\bibitem{jacak-royal}
J.~Jacak and L.~Jacak.
\newblock Explanation of nu=-1/2 fractional quantum hall state in bilayer
  graphene.
\newblock {\em Proc. R. Soc. A}, 472:20150330, 2016.

\bibitem{d6}
A.~Kumar, G.~A. Cs\'athy, M.~J. Manfra, L.~N. Pfeiffer, and K.~W. West.
\newblock Nonconventional odd-denominator fractional quantum hall states in the
  second landau level.
\newblock {\em Phys. Rev. Lett.}, 105:246808, 2010.

\bibitem{d7}
E.~Kleinbaum, A.~Kumar, L.~N. Pfeiffer, K.~W. West, and G.~A. Cs\'athy.
\newblock Gap reversal at filling factors $3+1/3$ and $3+1/5$: Towards novel
  topological order in the fractional quantum hall regime.
\newblock {\em Phys. Rev. Lett.}, 114:076801, 2015.

\bibitem{jac-zhetf-hll}
J.~Jacak and L.~Jacak.
\newblock The commensurability condition and fractional quantum hall effect
  hierarchy in higher landau levels.
\newblock {\em JETP Letters}, 102:19, 2015.

\end{thebibliography}
\end{document}